\documentclass[10pt,showpacs]{revtex4}
    \usepackage[dvips]{graphicx}

\begin{document} 

\title{Creation of macroscopic superposition states from arrays of Bose-Einstein condensates}
\author{J. A. Dunningham$^{1}$, K. Burnett$^{2}$, R. Roth$^{3}$, and W. D. Phillips$^{4}$}
\affiliation{${}^{1}$School of Physics and Astronomy, University of Leeds, Leeds LS2 9JT, UK 
\\
${}^{2}$Clarendon Laboratory, Department of Physics, University of Oxford, Oxford OX1 3PU, UK
\\
${}^{3}$Institut f\"ur Kernphysik, Technische Universit\"at Darmstadt, 64289 Darmstadt, Germany
\\
${}^{4}$National Institute of Standards and Technology, Gaithersburg, Maryland 20899}

\begin{abstract}
We consider how macroscopic quantum superpositions may be created from arrays of Bose-Einstein condensates. We study a system of 
three condensates in Fock states, 
all with the same number of atoms and show that this has the form of a highly entangled superposition of different quasi-momenta. 
We then show how, by partially releasing these condensates and detecting an interference pattern where they overlap, 
it is possible to create a macroscopic superposition of different relative phases for the remaining portions of the condensates. 
We discuss methods for confirming these superpositions.
 
\end{abstract}

\maketitle

Quantum mechanics allows objects to exist in a coherent superposition of different states. This does not depend on the size of the system and means that it should be possible to create superpositions of macroscopically
distinct states. Schr\"{o}dinger first put forward this idea in his well-known thought experiment entangling the 
fate of a cat with the state of a radioactively decaying nucleus \cite{schrodinger}. In this system, the cat is in a 
superposition of being alive and dead depending on the state of the nucleus. This outcome 
is at odds with our classical perception of the physical 
world and has generated a great deal of discussion about whether macroscopic superposition states can be observed.

Under conditions that carefully avoid decoherence \cite{Zurek}, 
it is possible to create superposition states in the laboratory. As time has gone on, these superpositions have been 
demonstrated in systems of increasing size including 
the spatial coordinates of a single atom \cite{monroe1996a}, the quantum phase of optical coherent states \cite{brune1996a}, 
the position of fullerene molecules \cite{Arndt1999a}, and the current in a superconductor \cite{friedman2000a}. 

As we shall see, Bose-Einstein condensates are a promising starting point for creating and studying cat states since they consist of a 
large number of particles all in the same quantum state and are sufficiently cold to enable 
quantum phase transitions. They can be considered to be analogous to lasers, which are useful for creating non-classical states of light. 
There have been a number of theoretical proposals 
for how macroscopic superpositions of condensates may be achieved \cite{beccat}.
In this paper, we focus on a system of condensates transferred to an optical lattice \cite{Orzel2001a,Greiner2002a}.
Although some authors reserve the term `cat state' for a macroscopic superposition entangled with a microscopic 
state in the sense of Schr{\"o}dinger's 
original thought experiment, we will use the term more generally in this paper to mean any superposition of 
macroscopically distinct states.

As our starting point we consider states where each site in the lattice 
has precisely the same number of atoms. We begin by showing 
that this state can be considered to be cat-like when viewed in quasi-momentum space. 
We then show how, by measuring interference patterns between overlapping clouds of this state, it is possible to 
create macroscopic superpositions of different relative phases between the condensates. 
The initial state may be created by a Mott insulator transition and it is helpful to begin by 
reviewing this transition for atomic condensates.

An array of condensates \cite{footnote1} can be created by applying, to a single trapped condensate, an optical lattice 
in the form of a standing wave. If this is done adiabatically in the sense that we remain in the ground state, 
the system can be described by the Bose-Hubbard Hamiltonian \cite{Jaksch1998a},
\begin{equation}
H=-J\sum_{<i,j>}a_{i}^{\dag}a_{j} + \frac{U}{2}\sum_{i} a_{i}^{\dag}a_{i}^{\dag}a_{i}a_{i}, \label{bosehubb}
\end{equation}
where $a_{i}$ is the annihilation operator for an atom at site $i$ and, in the first term, the summation 
is taken over nearest neighbors. The strength of the tunneling between sites,
$J$, can be adjusted in
experiments by changing the intensity of the standing wave, thus altering the potential barrier between sites. The interaction
strength between atoms, $U$, is at best only weakly dependent on the potential, but can be controlled by using Feshbach resonances. As $U/J$ is
increased, the atom number fluctuations at each site are progressively reduced \cite{Jaksch1998a} as has been experimentally demonstrated \cite{Orzel2001a}. 
For sufficiently large values of $U/J$, a quantum phase transition to the Mott insulator state has been predicted \cite{Jaksch1998a} and observed \cite{Greiner2002a}. 
In the limit $U/J \to\infty$, 
each site has precisely the same number of atoms assuming commensurability, i.e. the ratio of atoms to sites is an integer.

We shall take this perfectly-squeezed $n$-tuple Fock state, where $n$ is the number of sites, as our starting point. The simplest system required for our scheme is $n=3$ and we shall consider this case in detail. Three-well arrays of condensates have already been demonstrated in the laboratory \cite{Boyer2006a}. Later in the paper, we will consider $n=4$ and comment on the applicability of our results to larger arrays of condensates. For $n=3$, we label the atom annihilation operators for each condensate $a$, $b$, and $c$, respectively and take each condensate 
to initially have $N$ atoms. The state of the system can then be written in the atom number basis as,
\begin{equation}
|\psi\rangle = |N,N,N\rangle_{abc}. \label{Nmode}
\end{equation}
This is an idealized version of the state created in Mott transition experiments and we will consider below the effect of deviations from this due to imperfect squeezing.

To begin with, we would like to consider the form of this state (\ref{Nmode}) when it is transformed to a quasi-momentum basis, i.e. a complete basis of states 
that have equal weightings of $a$, $b$, and $c$ and a linearly varying phase across them.
The annihilation operators, $\{\alpha_{\xi},\beta_{\xi},\gamma_{\xi}\}$, corresponding to a general orthonormal quasi-momentum basis for three lattice sites are 
\begin{eqnarray}
\alpha_{\xi} &\equiv & \frac{1}{\sqrt{3}}\left(a + \mbox{e}^{-i\xi}b + \mbox{e}^{-i2\xi}c\right) \label{defalpha}\\
\beta_{\xi} &\equiv &  \frac{1}{\sqrt{3}}\left(a + \mbox{e}^{-i(\xi-2\pi/3)}b + \mbox{e}^{-i(2\xi-4\pi/3)}c\right) \label{defbeta}\\
\gamma_{\xi} &\equiv & \frac{1}{\sqrt{3}}\left(a + \mbox{e}^{-i(\xi+2\pi/3)}b + \mbox{e}^{-i(2\xi+4\pi/3)}c\right), \label{defgamma}
\end{eqnarray}
where the angle $\xi$ can take any value.
This basis also corresponds to the
outputs from six-port beam splitters (`tritters') if $a$, $b$, and $c$ are the inputs \cite{Walker1987a}.  These have been experimentally
realized for photons \cite{Mattle1995a}.

In this new basis, the initial state $|\psi\rangle$ is given by,
\begin{eqnarray}
|\psi\rangle &=& \frac{1}{\sqrt{N!^{3}}}\left(a^{\dag}b^{\dag}c^{\dag}\right)^{N}|0\rangle_{abc} \nonumber \\ 
&=& \frac{1}{\sqrt{(3^{N}N!)^{3}}}\left[\alpha_{\xi}^{\dag}{}^{3}+
\beta_{\xi}^{\dag}{}^{3}+\gamma_{\xi}^{\dag}{}^{3}
-3\alpha_{\xi}^{\dag}\beta_{\xi}^{\dag}\gamma_{\xi}^{\dag}\right]^{N}|0\rangle_{\alpha\beta\gamma}, \label{alphavac}
\end{eqnarray}
where $|0\rangle_{abc}$ and $|0\rangle_{\alpha\beta\gamma}$ denote the vacuum states in the 
two bases.
The state (\ref{alphavac}) can be written in the form
\begin{equation}
|\psi\rangle = \sum_{m=0}^{3N}\sum_{n=0}^{3N-m}f(m,n)|m,n,3N-m-n
\rangle_{\alpha\beta\gamma},
\end{equation}
where $f(m,n)$ are the coefficients found by operating on the vacuum state in (\ref{alphavac}).
The probability of finding $N_{\alpha}$ atoms in mode $\alpha$,
$N_{\beta}$ atoms in mode $\beta$ and hence $3N-N_{\alpha}-N_{\beta}$ atoms in
mode $\gamma$, is then
\begin{equation}
P(N_{\alpha},N_{\beta})=\left| f(N_{\alpha},N_{\beta})\right|^{2}.
\label{probdist}
\end{equation}

\begin{figure}[t]
\includegraphics[width=12.0cm]{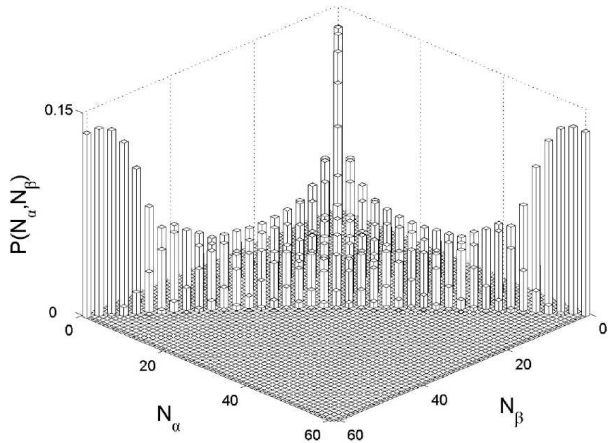}
\caption{The probability distribution given by (\ref{probdist}) for the number of atoms in the 
quasi-momentum modes $\alpha$ and $\beta$ for a state of the form $|\psi\rangle =|N,N,N\rangle_{abc}$ where $3N=60$.}
\end{figure}

This probability distribution is plotted in Figure~1 for $3N=60$ and is independent of the value of $\xi$ chosen in the mode decomposition. 
This state is analogous to that formed by passing Fock state pairs through a 50:50 beam splitter \cite{Holland}. 
It could be considered to have a `Schr\"{o}dinger cat-like' form in that the atoms tend to cluster around the corners, 
which correspond to all the atoms having the same quasi-momentum.
A `true cat' in the sense that we use in this paper would consist of a coherent superposition of all the atoms being in $\alpha$ 
and all in $\beta$ and all in $\gamma$. In the other extreme, if our state was $|\psi\rangle = |N,0,0\rangle_{abc}$, this would 
give $|\psi\rangle \propto (\alpha^{\dag}_{\xi}+\beta^{\dag}_{\xi}+\gamma^{\dag}_{\xi})^{N}|0,0,0\rangle_{\alpha\beta\gamma}$ 
in the quasi-momentum basis, i.e. it would be a microscopic superposition of different quasi-momenta. The state given by (\ref{alphavac}) 
and depicted in Figure~1 falls between these two extremes and we will call it a three-cornered `hat state'.

So far we have studied only the pure number states that emerge in the limit of 
strong repulsive interactions and weak tunneling. It is interesting to consider the effect of varying the 
interaction strength.
By analyzing the ground state that results from numerically diagonalizing the 
Bose-Hubbard Hamiltonian (\ref{bosehubb}), we can investigate how the distinct hat state structure evolves as a function of $U/J$. 

\begin{figure}[b]
\includegraphics[width=15cm]{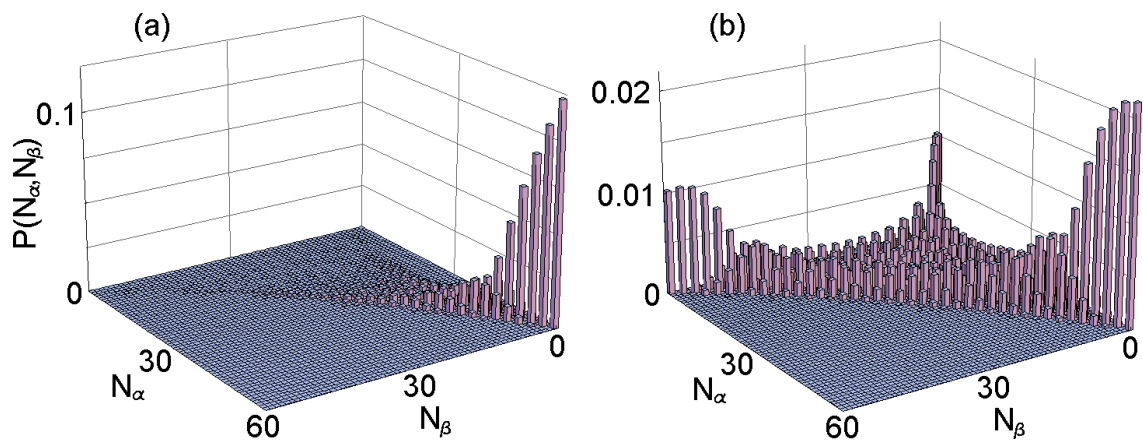}
\caption{Probability distribution (\ref{probdist}) for the ground state of the Bose-Hubbard 
Hamiltonian with $3$ sites and $60$ particles for (a) $U/J=50$ and (b) $U/J=500$.}
\end{figure}

Figure~2(a) shows the probability distribution (\ref{probdist}) for the ground state
of the Bose-Hubbard Hamiltonian with $U/J=50$. The interactions lead to a
depletion of the condensate mode $\alpha_{\xi}$, i.e. the ground state contains
admixtures of states with a few particles in modes $\beta_{\xi}$ and $\gamma_{\xi}$. 
If the ground state were a single pure condensate, as for $U/J=0$, Figure~2 would show a single peak at $(N_{\alpha},N_{\beta})=(60,0)$.
As one approaches the $n$-tuple Fock state regime, the distinct structure of Figure~1 
starts to emerge. For $U/J=500$, for example, the superposition of
macroscopically occupied modes becomes clear (see Figure~2(b)).
The symmetric three-cornered hat structure emerges only for the triple Fock state, which is to say in the limit of highly number squeezed systems.

States very close to our $n$-tuple Fock states have been experimentally created in large lattices \cite{Greiner2002a} and it would be 
interesting to see how they could be used. They may be well-suited to high precision measurement schemes in 
analogy with their two-site counterparts \cite{Holland,dunningham2002a}. 
In order to optimize proposed measurement schemes, we would like to start with a system with a large number of atoms in each lattice site 
since the phase resolution scales as $1/(d\sqrt{n})$, where $n$ is the number of sites and $d$ is the number of particles per site \cite{Vourdas2005a}. 
It is difficult to achieve the Mott transition in this regime \cite{Fisher}. One possibility may be to first perform 
the Mott transition in three dimensions with only a small number of atoms per site. 
The confining potential in two orthogonal directions could then be lowered sufficiently slowly that the system remains in the ground 
state of the evolving Hamiltonian. This would leave us with a one-dimensional array of condensates 
with the same (large) number of atoms in each site if the lattice had a cubic symmetry. The influence of different symmetries is an 
interesting question that would need to be addressed in any experimental implementation of this scheme.

We have seen how a cat-like state emerges in the quasi-momentum basis as the ratio $U/J$ is increased. Now we would like to turn our attention to how a `true' cat state can be created from this resource.  For simplicity, we will consider only the case that we begin with a perfectly squeezed state.  We shall show that, by partially releasing a triple Fock state (\ref{Nmode}) and detecting an interference pattern between the 
expanding atomic clouds, a cat state can be created in the relative phase of the remaining condensates. 
This is an extension of work that showed that a single relative phase is created between a pair of condensates when atoms are detected where they overlap \cite{Javanainen1996a, Naraschewski1996a,Castin1997a}.
The system we consider consists of three trapped condensates each initially containing $N$ atoms and spaced from each other 
by a distance, $d$.
At time $t=0$, the interactions are switched off and the condensates are partially released from the trapping potential and allowed to expand and overlap. 
This release could be achieved, for example, by coherently transferring part of the population to an untrapped state.
The overlapping atoms interfere and are detected at a distance $R$ from the lattice (see Figure~3).

The annihilation operator, ${\Omega}$, 
that corresponds to detecting an atom at angle $\theta$ and distance $R\gg d$ at time $t>0$ is,
\begin{equation}
{\Omega} = \frac{1}{\sqrt{3}}\left( {a}\,e^{-iE_{a}t/\hbar} + {b}\,e^{-iE_{b}t/\hbar} + {c}\,e^{-iE_{c}t/\hbar}\right), 
\label{etaeta}
\end{equation}
where $E_{a}$, $E_{b}$ and $E_{c}$ are respectively the energies required for atoms from $a$, $b$, 
and $c$ to have the correct speed to reach the screen at $\theta$ in time $t$. These kinetic energies, $E_{j}$, correspond to de Broglie wavelengths, 
$\lambda_{j}$, and can be rewritten as,
$E_{j}=mR_{j}^{2}/(2t^{2})$ for $j\in\{ a,b,c\}$, 
where $R_{a}$, $R_{b}$, and $R_{c}$ 
are the distances travelled by an atom from $a$, $b$, and $c$ to the point of detection and $m$ 
is the mass of an atom. 
For angles such that $\sin\theta \gg d/R$, we can write $(R_{a},R_{b},R_{c})=(R-d\sin\theta, R, R+d\sin\theta)$.
However, we need to ensure that we detect atoms at sufficiently small angles to correspond to the scale of 
variation for the interference fringes, 
i.e. $\sin\theta\sim \lambda/d$, where $\lambda$ is the typical de Broglie wavelength of the atoms. This gives us a 
condition for the geometry of the set-up, $\lambda/d \gg d/R$, i.e. $\lambda R\gg d^{2}$, which is the usual 
far-field condition guaranteeing that the size of the diffraction pattern is large compared with the source size.
Substituting into (\ref{etaeta}), we obtain,
\begin{equation}
{\Omega} = \frac{1}{\sqrt{3}}\left( {a} + {b}\, e^{-i\eta\sin\theta} + {c}\, e^{-2i\eta\sin\theta} \right), 
\label{etafinal}
\end{equation}
where $\eta = mRd/\hbar t$ and we have ignored any overall phase. 
Since $\eta$ depends on $t$, if many particles are detected as they arrive at a screen at various times, we get a range of values of 
$\eta$ and the fringes will tend to wash out. To avoid this, we can take a `snap-shot' of the spatial distribution of the overlapping 
clouds at a given time. The time we need to wait in order to satisfy the far-field condition is $t \gg md^2/h$.

\begin{figure}[t]
\includegraphics[width=9cm]{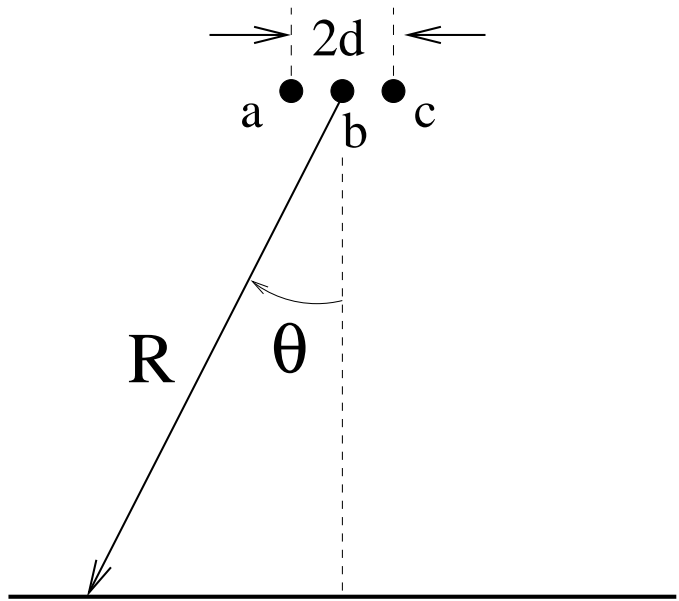}
\caption{General scheme for generating cats in relative phase space by observing interference between 
three condensates. In the far field $\lambda R\gg d^{2}$.}
\end{figure}

For simplicity, we will consider that atoms are detected at small angles and so the replacement 
$\sin\theta \approx \theta$ can be made in (\ref{etafinal}),
\begin{equation}
{\Omega} = \frac{1}{\sqrt{3}}\left( {a} + {b}\, e^{-i\eta\theta} + {c}\, e^{-2i\eta\theta} \right). \label{etafinal2}
\end{equation}

The emergence of cat states can be understood in terms of the information that the measurement process yields.
We first consider the case of initial 
coherent states with well-defined phases \cite{coherent},
\begin{equation}
|\psi\rangle = |\chi e^{i\phi_{a}},\chi e ^{i\phi_{b}}, \chi e^{i\phi_{c}}\rangle, \label{cohstate}
\end{equation}
where $|\chi|^{2}=N$, i.e. the mean number of particles in each coherent state is $N$. 
For large $N$, the probability density of atoms that would be observed if we were to take a `snap-shot' of the overlapping clouds at 
a time when most atoms satisfy the far-field condition, is
\begin{equation}
P(\theta, \eta) = \frac{1}{2\pi}\left( 1+ \frac{2}{3} F(\eta)\left[ \cos(\eta\theta-\phi_{ba}) + \cos(\eta\theta-\phi_{cb})+\cos(2\eta\theta-\phi_{ba}-\phi_{cb})\right]\right), \label{intprob}
\end{equation}
where $\phi_{ba}\equiv \phi_{b}-\phi_{a}$, $\phi_{cb}\equiv \phi_{c}-\phi_{b}$, and $F(\eta)$ is the normalized 
probability distribution for $\eta$.
We see that, independent of the form of $F(\eta)$, this interference pattern is identical under exchange of $\phi_{ba}$ 
and $\phi_{cb}$. This means that 
the measurement of such an interference pattern does not unambiguously establish the phase between the remaining condensates. Thus, the
initial triple Fock state, which is an equal superposition 
of all relative phases evolves, after partial detection, into a macroscopic superposition of two sets of relative phases. 
This is the principle behind the cat generation scheme. A similar calculation for four BECs, $a$, $b$, $c$, and $d$, shows 
that the distribution of particles detected is invariant under the exchange of the phases $\phi_{ba}$ and $\phi_{dc}$. This phase 
ambiguity suggests that cat states may also be created by this technique for arrays with $n=4$. It would be interesting to further investigate 
how cats emerge for $n>3$.
We shall now check the result for $n=3$ by direct numerical calculation.

\begin{figure}[b]
\includegraphics[width=10.0cm]{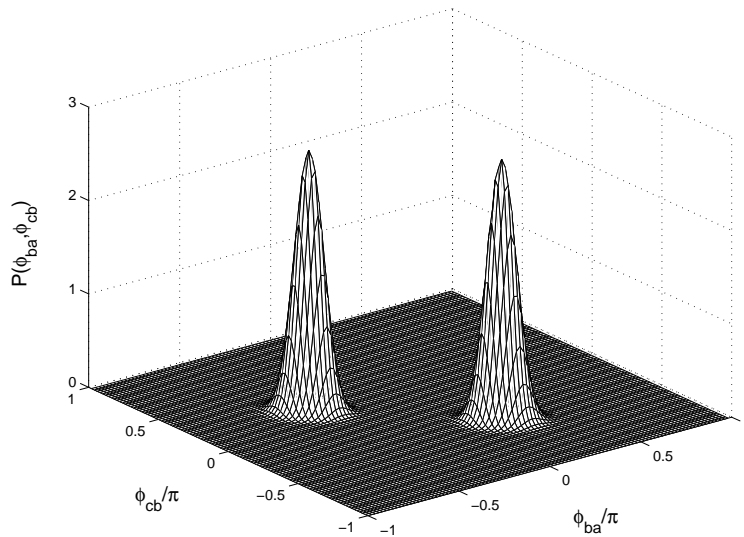}
\caption{Example of the joint relative phase probability distribution for a system of three
condensates (each initially in Fock states with 100 atoms) after 100 atoms have been detected in
the far field.} \label{cohcat}
\end{figure}

With the first detection of an atom at $\theta$, operator (\ref{etafinal2}) acts on state (\ref{Nmode}) to give,
\begin{equation}
|\psi\rangle = \frac{1}{\sqrt{3}}\left(|N-1,N,N\rangle + e^{-i\eta\theta}|N,N-1,N\rangle  + e^{-2i\eta\theta}|N,N,N-1\rangle\right). \label{detectone}
\end{equation}
The probability distribution for the relative phases, $\phi_{ba}$ and $\phi_{cb}$, of the remaining condensates is given by 
\begin{equation}
P(\phi_{ba},\phi_{cb}|\theta) = \left|\langle \phi_{ba},\phi_{cb}|\psi\rangle \right|^{2}, \label{relphaseprob}
\end{equation}
where $|\phi_{ba},\phi_{cb}\rangle$ is a three-mode generalization of the Pegg-Barnett relative phase state \cite {Barnett1990a} 
of the form,
\begin{equation}
|\phi_{ba},\phi_{cb}\rangle =
\sum_{p,q,r}e^{ip\chi}e^{iq(\chi+\phi_{ba})}e^{ir(\chi+\phi_{ba}+\phi_{cb})}|p,q,r\rangle, \label{peggstate}
\end{equation}
$|p,q,r\rangle$ represents a triple Fock state with $p$, $q$, and $r$ atoms in modes $a$, $b$, and $c$ respectively, 
and $\chi$ can take any value. Substituting (\ref{detectone}) and (\ref{peggstate}) into (\ref{relphaseprob}) we obtain,
\begin{equation}
P(\phi_{ba},\phi_{cb}|\theta) = 1 + \frac{2}{3}\left[ \cos(\eta\theta-\phi_{ba})+\cos(\eta\theta-\phi_{cb}) 
+\cos(2\eta\theta-\phi_{ba}-\phi_{cb})\right]. \label{P1}
\end{equation}
As predicted, this phase distribution is symmetric under exchange of $\phi_{ba}$ and $\phi_{cb}$. 
If the distribution is peaked at $(\phi_{ba},\phi_{cb})=(\Phi_{1},\Phi_{2})$, it will also be peaked 
at the converse $(\phi_{ba},\phi_{cb})=(\Phi_{2},\Phi_{1})$.
In other words, the state is a superposition of these two results and so a cat state begins to emerge.
This symmetry property can be shown to hold for subsequent detections, however the combinatorials 
get complicated very quickly and so it is much easier to numerically simulate the detection process.

Our system for this simulation initially consists of three condensates each in a Fock
state with 100 atoms. The condensates are partially released from the traps and detections are
made on a screen in the far field. 
In our simulations, we take the detected atoms to have a range of values of $\eta$ given by the  
distribution, $F(\eta)$. 
The interference pattern observed on the screen has the form of (\ref{intprob}) and so it is 
possible to determine the relative phases of the condensates. However, there is still an ambiguity as to which 
condensates the phases should be assigned.
In Figure~4 we have plotted an example of the results for the relative phase probability 
distribution, $P(\phi_{ba},\phi_{cb})$, of the state after 100 atoms have been detected. 
As expected, the state of the remaining portions of the condensates has evolved into a 
Schr\"{o}dinger cat in their relative phases.
The state shown here is a superposition of $\phi_{ba}\approx -0.4\pi$ and $\phi_{cb}\approx0.2\pi$ 
with the reverse, $\phi_{ba}\approx 0.2\pi$ and $\phi_{cb}\approx -0.4\pi$.
The positions of the peaks shown in Figure~4 differ from run to run.

\begin{figure}[t]
\includegraphics[width=10cm]{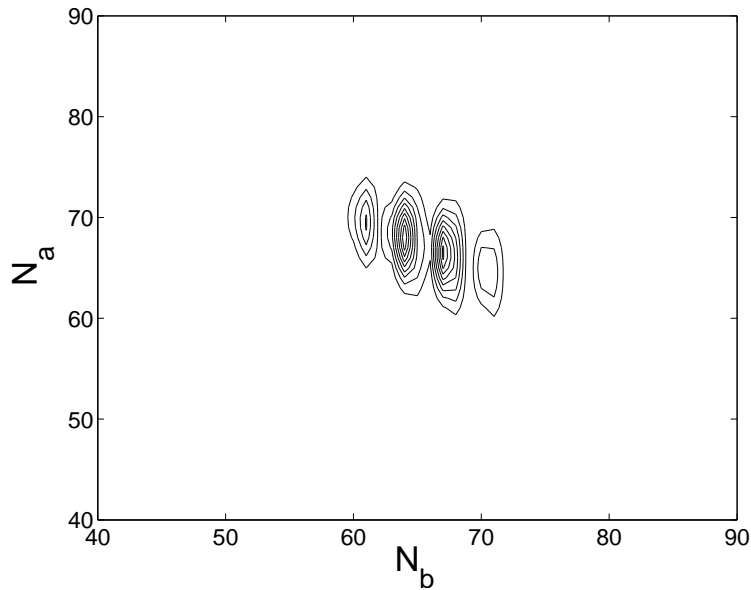}
\caption{Contour plot of the number distribution of atoms in $a$ and $b$ corresponding 
to the state shown in Fig.~\ref{cohcat}.}
\end{figure}

This demonstrates how it may be possible to create large Schr\"odinger cat states by simply measuring interference 
patterns from an array of $n$-tuple Fock state condensates. A particularly pleasing feature is that every trial 
successfully creates a cat state and that the interference pattern itself tells us what the two superposed 
phases are. This means that we always create a cat of a known form, but with unpredictable phases.

During this process, there is an interplay between the detections, which tend to create relative phases between the 
condensates, and interactions, which tend to wash the phases out. It has been shown that a balance is reached between these 
two effects \cite{Kohler2001a} and so interactions between the atoms will not prevent cat states from forming. When detections are 
no longer made, the relative phases will undergo collapses and revivals \cite{Wright1996a}. The effect of the interactions could be reduced by making use of Feshbach resonances to tune the scattering length between atoms \cite{Cornish2000a}.

The next challenge is to devise methods for seeing these cats directly and confirming the coherence between the two parts. 
For the system considered here, this could be achieved simply by looking for interference fringes in number space. 
We can see this by considering a state of the form depicted in Figure~4 for which $(\phi_{ba},\phi_{cb})=(\Phi_{1},\Phi_{2})$ 
and the converse. After many detections, this state can approximately be written as,
\begin{equation}
|\psi\rangle = \frac{1}{\sqrt{2}}\sum_{p,q} C_{p,q}\left(\mbox{e}^{-ip\Phi_{1}}\mbox{e}^{i(2N-p-q)\Phi_{2}}
+ \mbox{e}^{i\delta}\mbox{e}^{-ip\Phi_{2}}\mbox{e}^{i(2N-p-q)\Phi_{1}}\right)|p,q,2N-p-q\rangle,
\end{equation}
where $C_{p,q}$ is the probability amplitude that there are $p$ atoms in $a$ and $q$ in $b$. There are a total of $2N$ 
particles in the system since $N$ were detected in creating the state. The additional phase $\delta$ between the terms depends on the particular detections made and so, although it is random, it can be known for a particular experimental run. The probability 
distribution for the number of atoms in $a$ and $b$ is then given by,
\begin{eqnarray}
P(N_{a},N_{b}) &=& |\langle N_{a},N_{b},2N-N_{a}-N_{b}|\psi\rangle|^{2} \nonumber \\
&=& |C_{N_{a},N_{b}}|^{2}\left( 1 + \cos[(2N-N_{b})(\Phi_{1}-\Phi_{2}) + \delta ]\right). \label{numdist}
\end{eqnarray}
This predicts fringes that vary with $N_{b}$ but not with $N_{a}$.
In Figure~5, we have plotted the 
number distribution for the cat state depicted in Figure~4 and resulting from the example simulation. 
This confirms that interference fringes of the form 
of (\ref{numdist}) are present.
In practice, this distribution would be obtained by an ensemble of measurements. We could create cat states similar to the one 
shown in Figure~4 by making measurements in the far field and, whenever a particular pair of relative phases is obtained, we 
could then directly measure the number of atoms in the remaining condensates. Repeating this procedure would enable us to build up 
the number distribution in Figure~5.
No fringes are predicted   
for states that are not a superposition of different phases and so an 
experimental observation of interference would provide strong evidence that a Schr{\"o}dinger cat state had been created.

This work was financially supported by the United
Kingdom EPSRC Advanced Fellowship GR/S99297/01, the Royal Society and Wolfson Foundation, NIST, ONR, and ARDA.


\begin{thebibliography}{10}
\bibitem{footnote1} We will use the term `condensate' throughout this paper to refer to atoms in the ground state of the trapping potential 
even though this may sometimes refer to only one or two atoms.
\bibitem{coherent} This approach also works when the total number of atoms in the system is fixed. We use a coherent state approach here to simplify the analysis.
\end{thebibliography}
\end{document}